\documentclass[showpacs,twocolumn,prb,floatfix,amssymb,aps,superscriptaddress]{revtex4-2}

\usepackage{graphicx}
\usepackage{dcolumn}
\usepackage{bm}
\usepackage{setspace}

\begin{document}

\title{Topological spin textures in chiral magnets on the honeycomb lattice with magnetic fields}

\author{Li-Peng Jin}
\email{11930771@mail.sustech.edu.cn}
\affiliation{ Department of Physics, Southern University of Science and Technology, Shenzhen 518005, China }

\author{Bin Xi}
\email{xibin@yzu.edu.cn}
\affiliation{ College of Physics Science and Technology, Yangzhou University, Yangzhou 225002, China }

\author{Yong-Jun Liu}
\affiliation{ College of Physics Science and Technology, Yangzhou University, Yangzhou 225002, China }

\date{\today}

\begin{abstract}
Topological spin textures, like skyrmions, have significant potential for spintronics applications. The main purpose of this work is to study further the topological spin textures on the discrete lattice with magnetic fields. In this work, we study a classical rotated Heisenberg model with Dzyaloshinskii-Moriya interaction, bond-dependent anisotropy and easy-axis anisotropy on the honeycomb lattice via Monte Carlo simulations. We mainly focus on phase diagrams with magnetic fields, especially on the non-trivial phases only existing with fields. The results demonstrate the emergence of field-induced incommensurate skyrmion superlattice including mixed skyrmion-bimeron states, ferromagnetic star, vortex and z-vortex superlattice. We systematically analyze structures of these topological spin textures through spin configurations, structure factors, topological charge and vector chirality. We hope that our results could be useful for the identification of topological phases experimentally.  
 
\end{abstract}

\maketitle

\section{Introduction}
Magnetic skyrmions \cite{Skyrme62,Bogdanov94} are representative topological spin textures, which have been observed or proposed due to four main mechanisms, such as magnetic dipolar interaction\cite{Lin73,Yu12}, Dzyaloshinskii-Moriya (DM) interaction\cite{Dzyaloshinsky58,Moriya60,M09,Yu10}, frustrated exchange interaction\cite{Okubo12} and four-spin exchange interactions\cite{Heinze11}. The DM interaction arises from the relativistic spin-orbit coupling (SOC), which has been extensively studied in continuous models, such as in chiral magnets\cite{Nagaosa13,Iwasaki13,Lin14,Tokunaga15,Fert17}. In addition, due to the experimental realization of synthetic SOC for ultra-cold atoms\cite{Lin11}, more attention has been paid to discrete lattice models, such as the extended Hubbard model on square lattice\cite{Cole12}. In the strong coupling limit, this model could derive the rotated Heisenberg model with DM interaction and bond-dependent anisotropy\cite{She92,Yildirim94,Banerjee14,Sun15,Zhao15,Xi17,Cai12}, which support magnetic textures such as spin spirals and vortex and skyrmion superlattice.

However, the magnetic fields also play an important role in the formation of topological spin textures\cite{Yu18,Wang21}. Motivated by these, we study further the rotated Heisenberg model on the honeycomb lattice via Monte Carlo simulations. We mainly focus on phase diagrams with out-of-plane magnetic fields, especially on the non-trivial phases only existing with fields. The results of zero fields are consistent with the previous work\cite{Cole12,Sun16}. On the contrast, the results with fields indicate the emergence of ferromagnetic star (FM star)\cite{Janssen16}, incommensurate skyrmion superlattice (IC-SkL) including mixed skyrmion-bimeron states\cite{Iakovlev18}, and z-vortex superlattice (Z-VL).  We demonstrate a distinct type of topological structure, bimeron that consists of two merons (half-skyrmions) separated by a spiral domain with zero topological charge, which could further turn into skyrmion superlattice upon increasing the magnetic fields. The FM star contains a large hexagon with the in-plane spin component rotating $2\pi$ in the core hexagon and six spins in the outer hexagon aligning in a parallel fashion. We characterize all these topological spin textures and give the structure factors for the identification of topological phases experimentally. 

The paper is organized as follows. In Sec.~\ref{sec-Model}, we start
from the bosonic model on the honeycomb lattice with a synthetic SOC. In the strong coupling limit, we obtain the rotated Heisenberg model with DM interaction, bond-dependent anisotropy and easy-axis anisotropy. We give details about the Monte Carlo simulation and how to characterize different spin configurations. In Sec.~\ref{sec-Ground}, we map out phase diagrams with/without fields and mainly focus on phase diagrams with fields, especially on the phases only existing with fields. In Sec.~\ref{sec-magnetic}, we describe the structures of topological spin textures and give the structure factors for the identification of these topological phases experimentally. In Sec.~\ref{sec-discuss}, we give our discussion and conclusion.

\section{Model and Method}\label{sec-Model}
The low-energy behavior of bosons on the honeycomb lattice with SOC could be described with the following Hamiltonian:
\begin{eqnarray}
	\hat{\mathcal{H}}_{\textrm{boson}} & = & -t \sum_{\langle i j\rangle}\left(\hat{c}^{\dagger}_{i\tau} \mathcal{R}_{i j} \hat{c}_{j\tau^\prime}+\mathrm{H.c.}\right) + \hat{\mathcal{H}}_U, \nonumber \\
	\mathcal{R}_{i j} & = & \exp \left[i \boldsymbol{A} \cdot\left(\boldsymbol{r}_i-\boldsymbol{r}_j\right)\right], \nonumber\\
	\hat{\mathcal{H}}_U & = &\frac{U}{2} \sum_{i\tau} \hat{n}_{i\tau} ( \hat{n}_{i\tau} - 1)+ U^{\prime} \sum_i \hat{n}_{i\uparrow} \hat{n}_{i\downarrow} ,
	\label{OriginH}
\end{eqnarray}
where the first term in $\hat{\mathcal{H}}_{\textrm{boson}}$ describes the hopping term between nearest-neighbor sites, with $t$ representing the hopping amplitude and $\hat{c}^\dagger_{i\tau}$ ($\hat{c}_{i\tau}$) is the creation (annihilation) operator at site $i$, with subscripts $\uparrow$ and $\downarrow$ representing two internal hyperfine states of bosons. The $\boldsymbol{A}=(\theta \sigma_{y},-\theta \sigma_{x},0)$ in matrix $\mathcal{R}_{i j}$ is a non-Abelian gauge field by the bosons, which is the well-known Rashba term. The second term $ \hat{\mathcal{H}}_U$ describes on-site interactions between bosons,
where $U$ is intracomponent interaction and $U^{\prime}$ is the intercomponent one, $\hat{n}_{i\tau}=\hat{c}^\dagger_{i\tau}\hat{c}_{i\tau}$ is the boson number operator with spin $\tau$ at site $i$.

At unit filling and in the strong coupling limit $U / t \gg 1$, the bosonic model can be approximated as a rotated Heisenberg model, the effective low-energy spin Hamiltonian given by:
\begin{eqnarray}
	\hat{\mathcal{H}}_{\text{spin}} & = & J\sum_{\langle i,j \rangle} \left\{  J_0 \bm{S}_i \cdot \bm{S}_j + \bm{D}_{i j}\cdot (\bm{S}_{i}\times \bm{S}_{j}) \right. \nonumber \\
	&  &+A_c \left[\bm{S}_{i} \cdot\left(\bm{\hat{e}}_z \times \bm{r}_{i j} \right)\right]\left[\bm{S}_{j} \cdot\left(\bm{\hat{e}}_z \times \bm{r}_{i j}\right)\right]  \} \nonumber\\
	& &+J\sum_{i} \Delta({S_i}^{z})^2, \nonumber \\
	J=&-&4 t^{2} / U,  \quad J_0=\cos 2\theta, \nonumber \\
    \bm{D}_{i j}&=&\sin 2\theta (\bm{\hat{e}}_z \times \bm{r}_{i j}),  A_c=2\sin^2 \theta ,
	\label{HEFF}
\end{eqnarray}
we set $|J|=1$ as the energy unit scale. The first term is the nearest-neighbor Heisenberg interaction. The second term is DM interaction, which is anisotropic antisymmetric interaction. The third term is bond-dependent anisotropy, which is symmetry-allowed even in the presence of inversion symmetry. 
The last term in $\hat{\mathcal{H}}_{\text{spin}} $ is easy-axis anisotropy $(\Delta>0)$, which is derived from intercomponent interaction $U^{\prime}$ between bosons.

\begin{figure} [ht!]
	\begin{center}
		\includegraphics[width=0.65\linewidth]{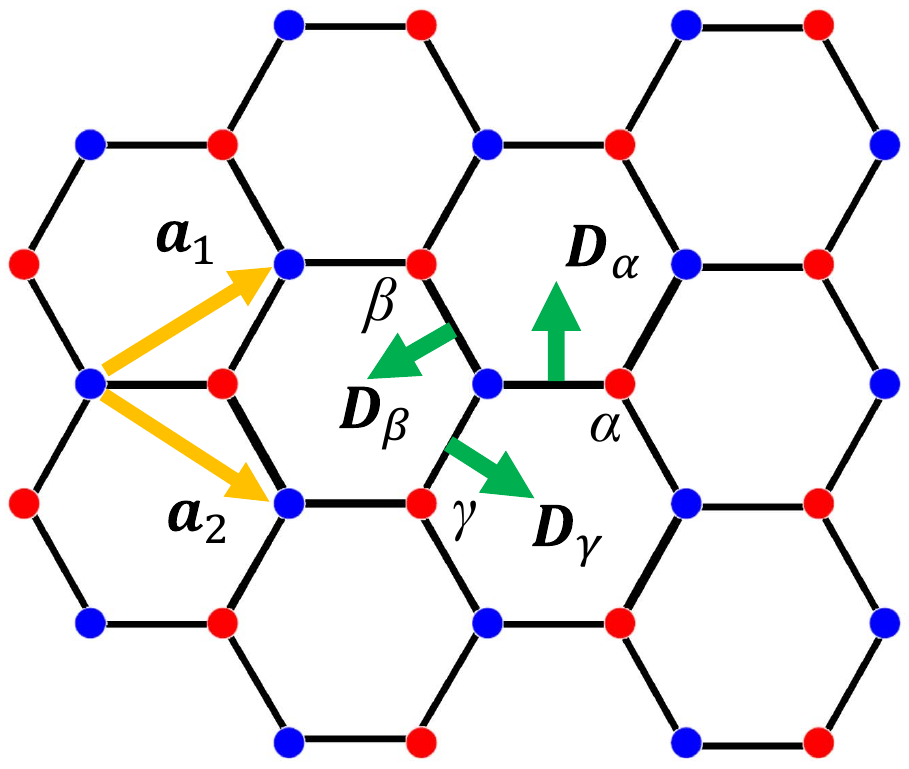}
	\end{center}
	\caption{   Sketch of the honeycomb lattice, with the primitive vectors $\bm{a}_{1}, \bm{a}_{2}$, and $\bm{D}_{\alpha}, \bm{D}_{\beta}, \bm{D}_{\gamma} $ corresponding to the DM vectors in $\alpha$, $\beta$ and $\gamma$ bonds, respectively. }
	\label{fig1}
\end{figure}

We study the honeycomb lattice with $L\times L$ ($L=60$) unit cells placed in the $x y$-plane, where each unit cell contains two sites colored with blue and red points shown in Fig.~\ref{fig1}, hence the system contains $2\times L\times L=7200$ sites. We set the lattice constant $a=1$ and the magnitude of the spins is fixed by the normalization condition $|\bm{S}_{i}|=1$.
The periodic boundary conditions are imposed. Due to the anisotropy of the honeycomb lattice, there are three types of $\bm{r}_{i j}$ pointing from blue site to red site:
$\bm{r}_{i j}^\alpha=(1,0)$, $\bm{r}_{i j}^\beta=(-1/2, \sqrt{3}/2)$ and $\bm{r}_{i j}^\gamma=(-1/2, -\sqrt{3}/2)$. The three types of DM vectors ($\bm{D}_{\alpha}, \bm{D}_{\beta}, \bm{D}_{\gamma}$) are shown with green arrows. We use the convention $\bm{D}_{i j}\cdot(\bm{S}_{i} \times \bm{S}_{j})$ in which the first spin in the cross product
$\bm{S}_{i}$ is always on the sublattice color with blue. The lattice is constructed from different sets of primitive vectors $(\bm{a}_{1}, \bm{a}_{2})$ shown with yellow arrows.

To reveal the true classical ground states of $\hat{\mathcal{H}}_{\text{spin}} $ in Eq.~(\ref{HEFF}), we use the paralleling-annealing Monte Carlo (MC) simulation\cite{Hukushima96} on 40 replicas, with temperature $T/|J|$ ranging from $0.001$ to $1.0$.
For each replica, we sample it with a combination of heat-bath and over-relaxation methods\cite{Miyatake86}. A whole MC step consists of a single heat-bath sweep and subsequent ten over-relaxation sweeps over the entire lattice.
We perform $2\times 10^5$ MC steps per replica, then we copy out the spin configuration from the lowest-T replica and sample it with a combination of zero-temperature heat-bath and over-relaxation method to obtain the ground states.
The zero-temperature heat-bath sampling is simply aligning the spins according to their local fields:
\begin{eqnarray}
	\bm{S}_i &=& \frac{\bm{h}_i^{loc}}{|\bm{h}_i^{loc}|} S,
\end{eqnarray}
with $\bm{h}_i^{loc}={\partial \hat{\mathcal{H}}_{\text{spin},i}}/{\partial \bm{S}_i} $.

The important observables that can be used for the identification of magnetic phases are the spin
structure factors given by:
\begin{eqnarray}
	\bm{F_k} &=& \frac{1}{N}\sum_{ij} \left \langle \bm{S}_i \cdot \bm{S}_j\right \rangle e^{i\bm{k}\cdot (\bm{r}_i-\bm{r}_j)}, 
	\label{Sk}
\end{eqnarray}
where $\bm{k}$ is the reciprocal space vector, $\bm{r}_i$ is the lattice vector at site $i$ and $N$ the number of sites.

Each skyrmion contributes $\pm 1$ to the topological charge (skyrmion number). To calculate the topological charge, we employ the definition on the lattice version introduced by Berg\cite{Berg81,Yin16}. The calculation of $Q$ starts
by triangulating the entire lattice and then counting the solid angle $\Omega_l$ for each triangle $l$ through all three spins $(\bm{S}_1, \bm{S}_2, \bm{S}_3)$ located at its vertices, and the topological charge can be obtained as $Q=\sum_{l}\Omega_l/4\pi$

We also calculate the vector chirality $\bm{\kappa}$ for each triangle:
\begin{eqnarray}
	\bm{\kappa}  = \frac{2}{3\sqrt{3}} \left( \bm{S}_1\times \bm{S}_2+\bm{S}_2\times \bm{S}_3 +\bm{S}_3\times \bm{S}_1 \right)
\end{eqnarray}
where three spins $\bm{S}_1$, $\bm{S}_2$ and $ \bm{S}_3$ are chosen anti-clockwise on the triangle.

\section{Results}\label{Results}
\subsection{Phase diagrams}\label{sec-Ground}
We obtain the classical ground-state phase diagram of $\hat{\mathcal{H}}_{\text{spin}}$ in Eq.~(\ref{HEFF}) via the MC simulation, with tuning the parameter $\theta$ and the easy-axis anisotropy $\Delta$. We map out phase diagrams with/without and get the following phases characterized by the spin structure factors $\bm{F_k}$. We mainly focus on phase diagram with magnetic fields, especially on the non-trivial phases only existing with fields.
To identify the different topological phases realized in the phase diagrams, we used the calculated topological charge and vector chirality to visualize magnetic configurations.

\subsubsection{phase diagram without magnetic fields}
\begin{figure} [ht!]
	\begin{center}
	\includegraphics[width=0.76\linewidth]{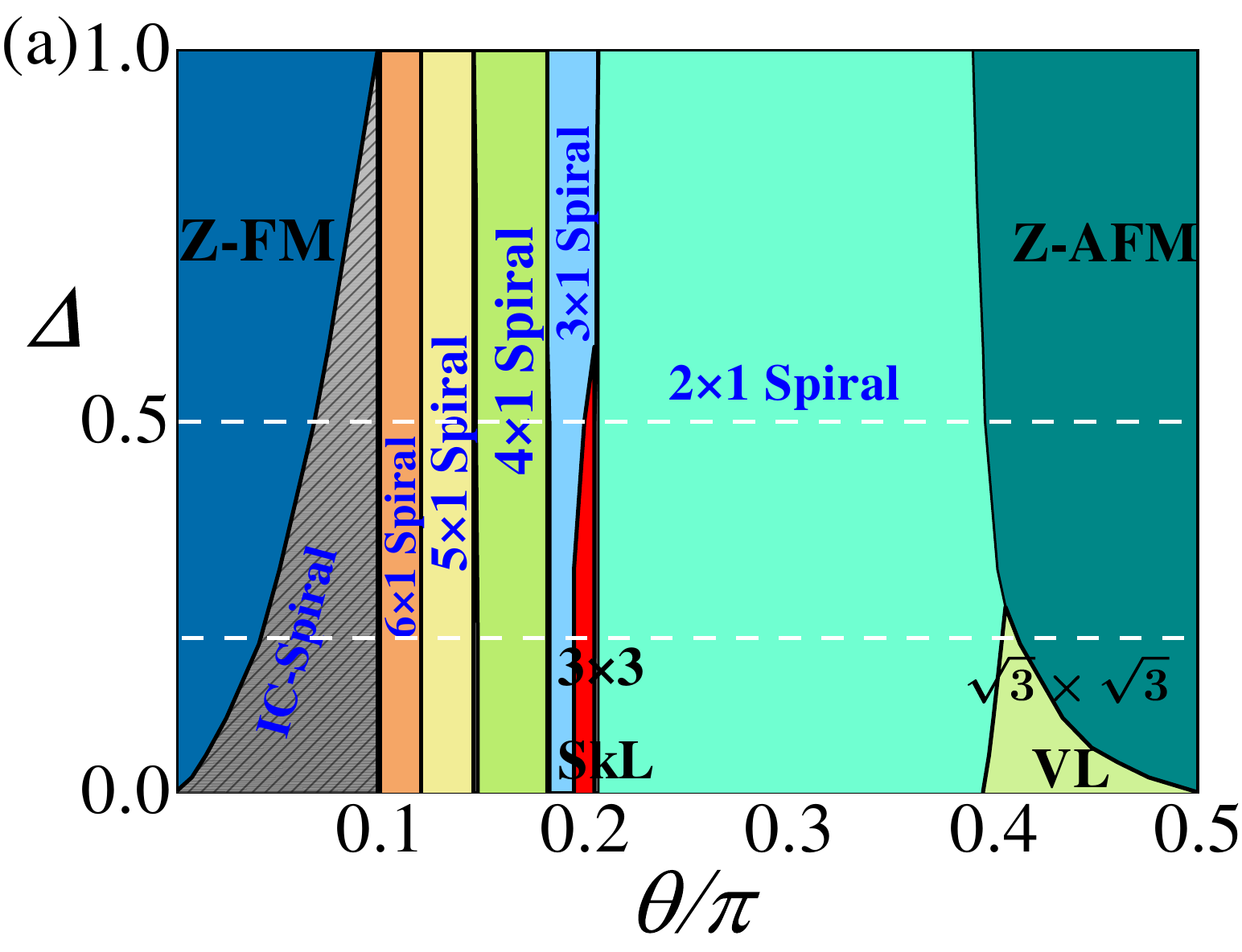}
	\includegraphics[width=0.75\linewidth]{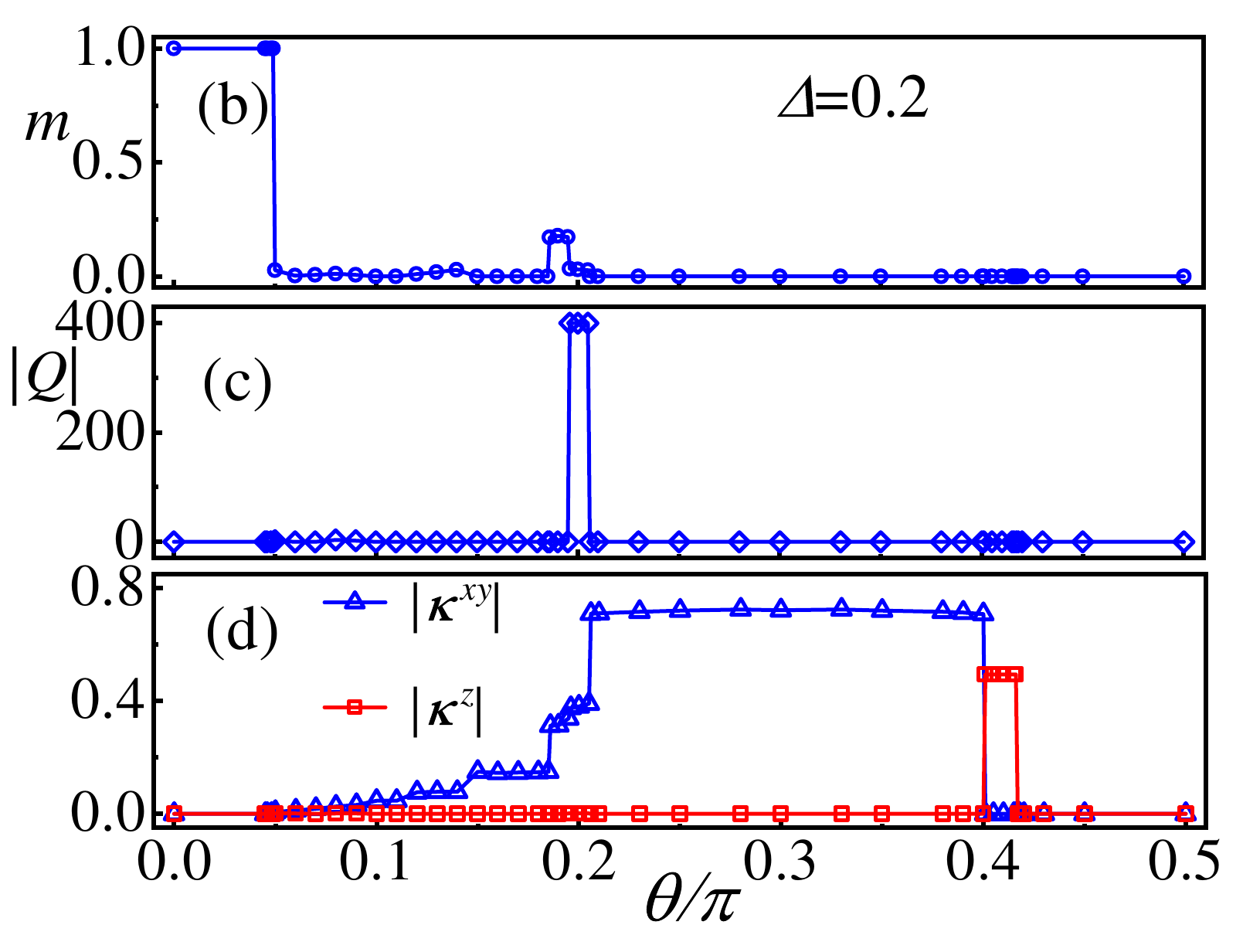}
	\end{center}
	\caption{  (a) phase diagram of the $\hat{\mathcal{H}}_{\text{spin}} $ in Eq.~(\ref{HEFF}) with tuning easy-axis anisotropy $\Delta$ and the parameter $\theta$. Spin configurations are abbreviated as described in the text. (b)-(d) show order parameter as a function of $\theta$ without magnetic fields at $\Delta=0.2$. (b) The average magnetic moment $m$. (c) The topological charge $|Q|$. (d) The chirality $\bm{\kappa}^{xy}$ and $\bm{\kappa}^{z}$. } 
	\label{zero}
\end{figure}

In Fig.~\ref{zero}(a), we plot the phase diagram without magnetic fields, which is consistent with the previous work on the square lattice, including ferromagnetic phase (Z-FM), antiferromagnetic phase (Z-AFM), spiral phases and two topological phases ($3 \times 3$ SkL and $\sqrt{3} \times \sqrt{3}$ VL). The $3 \times 3$ SkL and $\sqrt{3} \times \sqrt{3}$ VL phases mean the $3 \times 3$ unit cell skyrmion superlattice and $\sqrt{3} \times \sqrt{3}$ unit cell vortex superlattice, respectively. The order parameters in FM, AFM, SkL, VL and spirals are given by average magnetic moment $m$ $(m=N^{-1} \sum_{i}{S_i}^{z})$, topological charge $|Q|$ and vector chirality $\bm{\kappa}$, respectively. In Fig.~\ref{zero}(b)-(d), we plot these order parameters as a function of $\theta$ at $\Delta=0.2$. 

\subsubsection{phase diagram with magnetic fields}
\begin{figure*} [ht!]
	\begin{center}
		\includegraphics[width=0.42\linewidth]{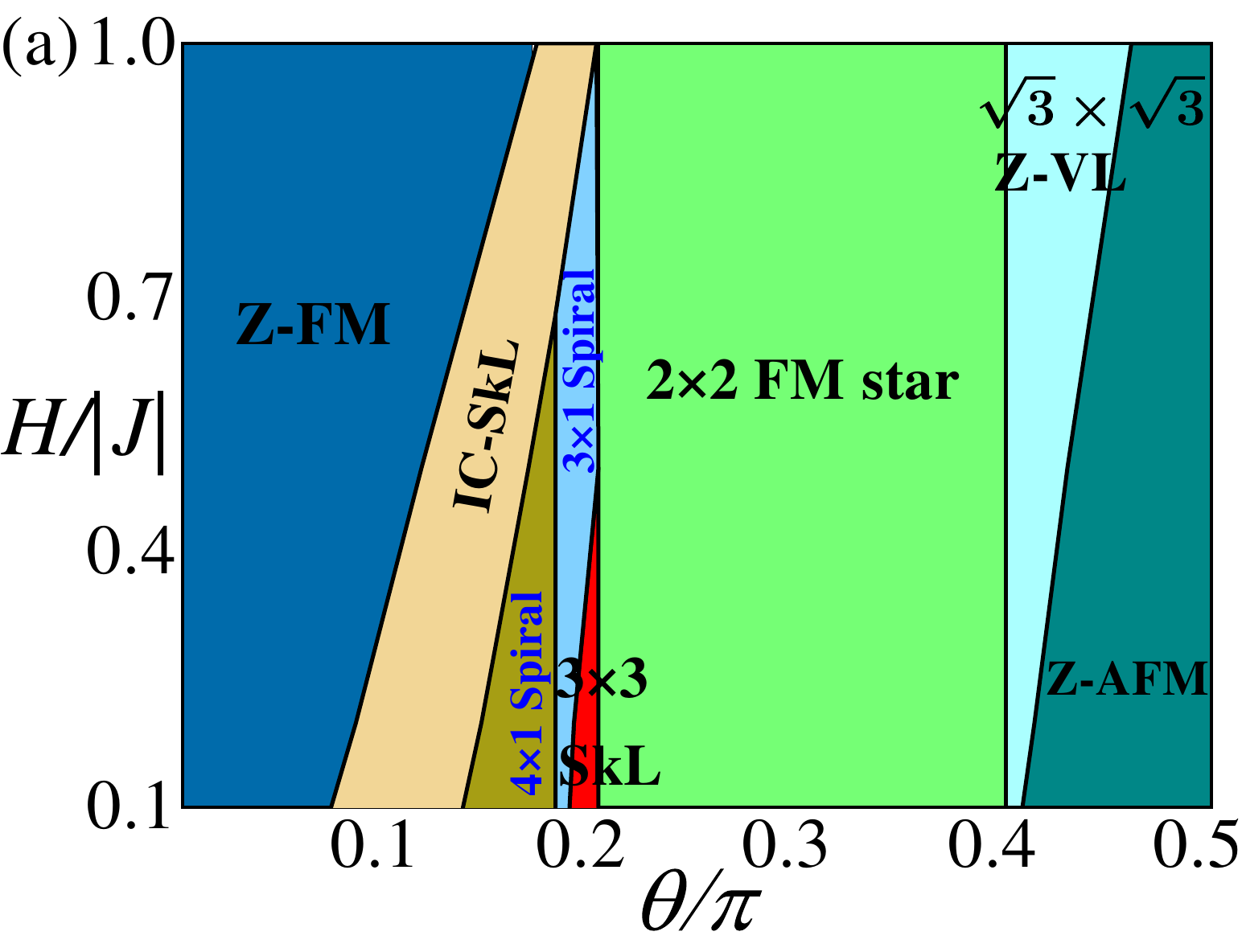}
		\includegraphics[width=0.42\linewidth]{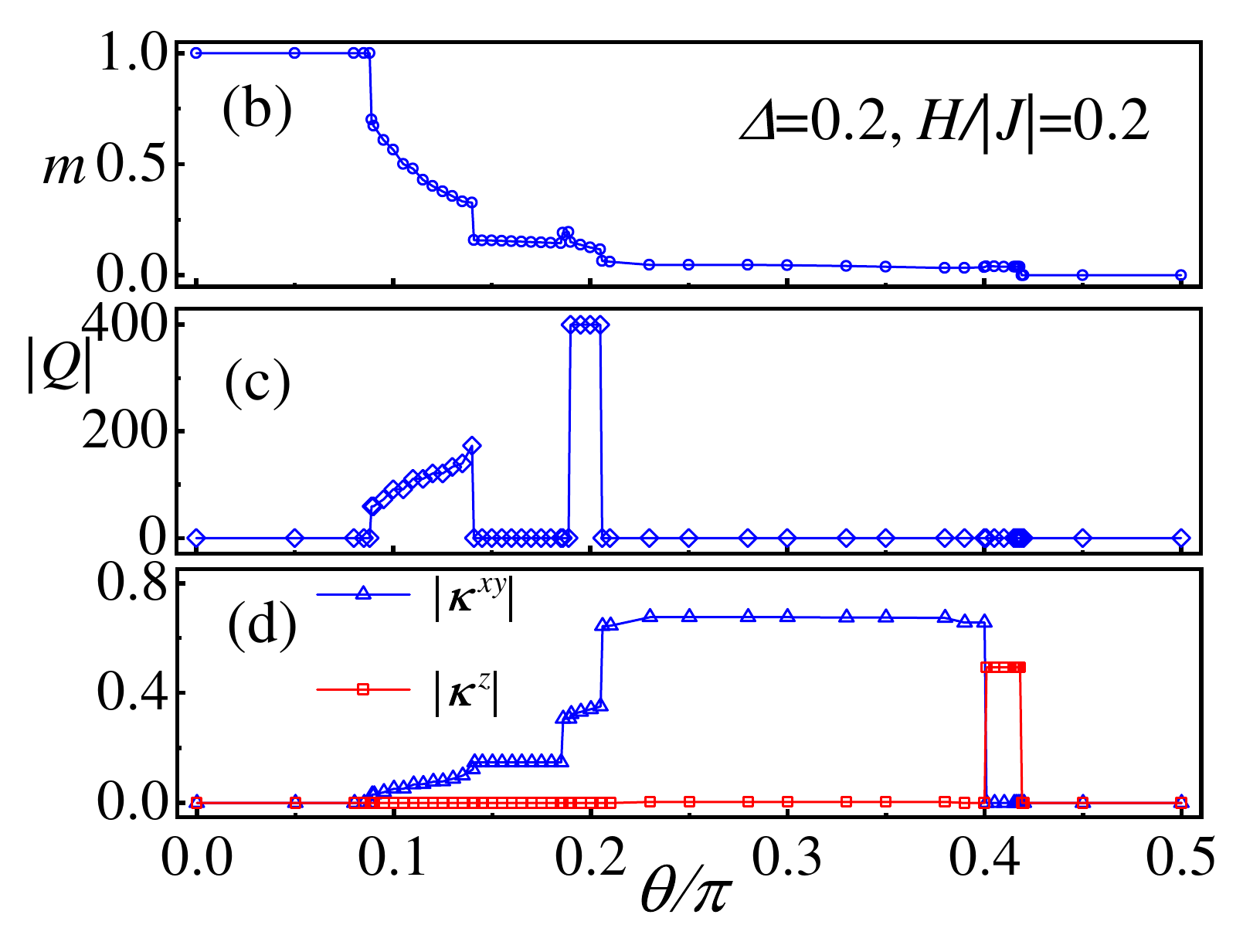}
		\includegraphics[width=0.42\linewidth]{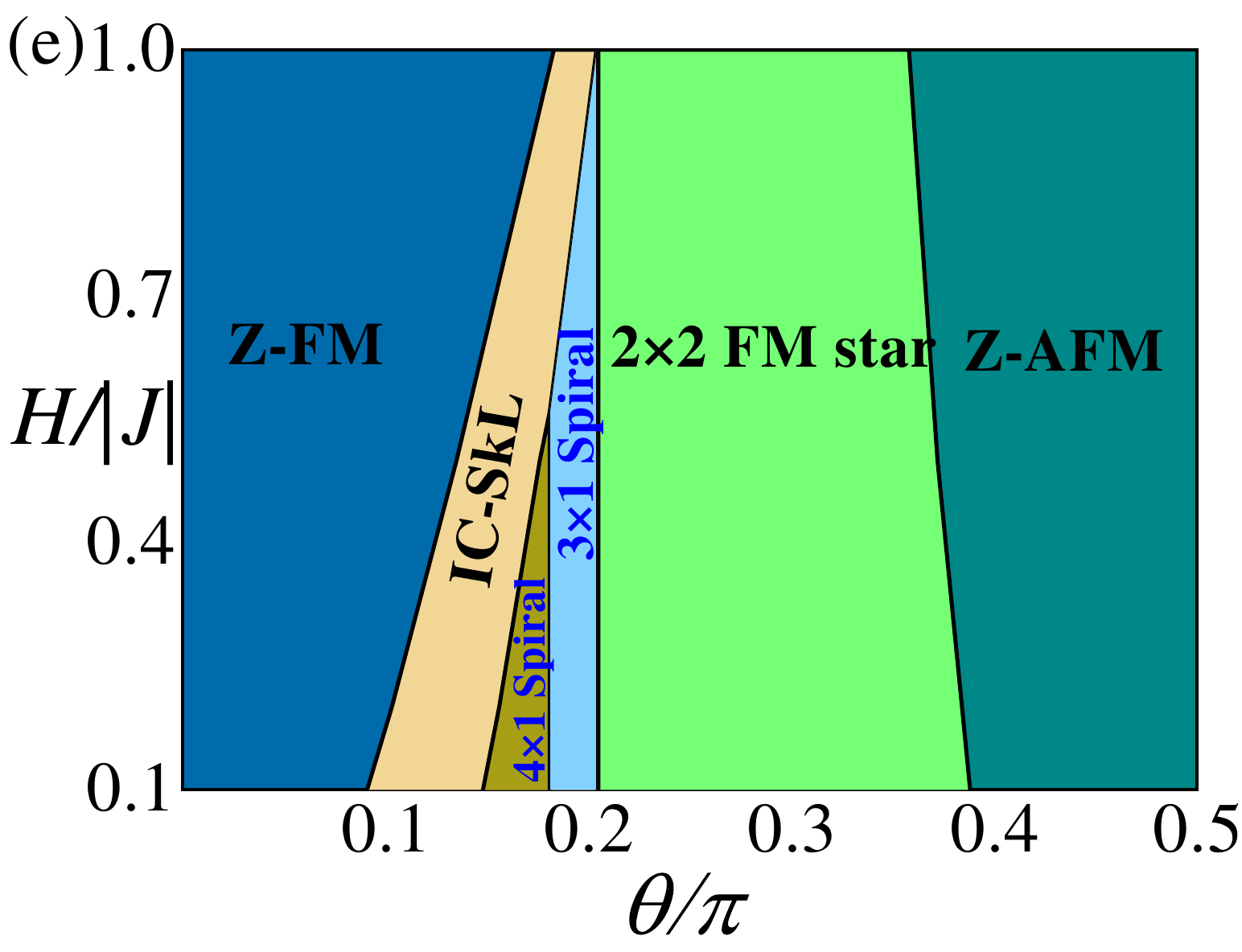}
		\includegraphics[width=0.42\linewidth]{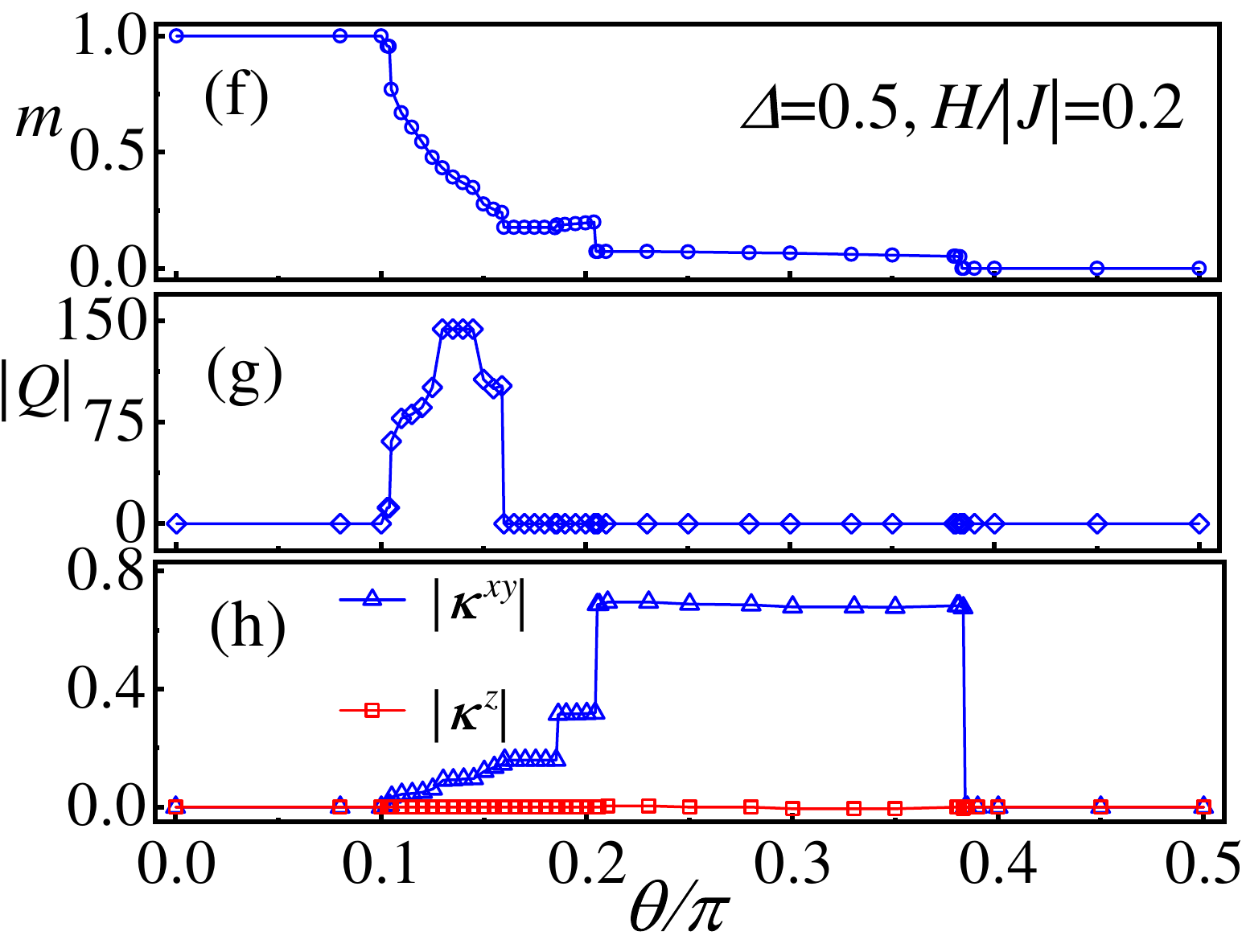}
	\end{center}
	\caption{  Phase diagram of the $\hat{\mathcal{H}}_{\text{spin}}$ with  out-of-plane magnetic fields, (a) and (e) corresponding to different easy-axis anisotropy $\Delta=0.2$ and $0.5$, marked with dashed lines in Fig.~\ref{zero}. (b)-(d) show order parameter as a function of $\theta$ at $\Delta=0.2$, $H/|J|=0.2$. (b) The average magnetic moment $m$. (c) The topological charge $|Q|$. (d) The chirality $\bm{\kappa}^{xy}$ and $\bm{\kappa}^{z}$. (f)-(h) show order parameter as a function of $\theta$ at $\Delta=0.5$, $H/|J|=0.2$. (f) The average magnetic moment $m$. (g) The topological charge $|Q|$. (h) The chirality $\bm{\kappa}^{xy}$ and $\bm{\kappa}^{z}$.}
	\label{field}
\end{figure*}
With applied out-of-plane external magnetic fields in the Hamiltonian of $\hat{\mathcal{H}}_{\text{spin}} $, we obtain two different phase diagrams corresponding to different easy-axis anisotropy ($\Delta=0.2$ and $0.5$), marked with dashed lines in Fig.~\ref{zero}(a). In Fig.~\ref{field}(a) and Fig.~\ref{field}(e), we map out the two phase diagrams with out-of-plane external magnetic fields.
Most of the phases in the phase diagrams of ground states are still stable when applying external magnetic fields. It is wondrous that three field-induced phases have emerged, viz IC-SkL phase, $2 \times 2$ FM star phase and $\sqrt{3} \times \sqrt{3}$ Z-VL phase, respectively. 
The order parameters in these phases are also given by average magnetic moment $m$, topological charge and chirality, respectively. In Fig.~\ref{field}, we plot these order parameters as a function of $\theta$ with different parameters.  Significantly, the topological charge $|Q|$ shown in Fig.~\ref{field}(c) and Fig.~\ref{field}(g) indicate that the number of skyrmions in IC-SkL phase varies with different parameters.

We briefly talk about the non-trivial phases in phase diagrams including ferromagnetic phase, antiferromagnetic phase and spiral phases. Z-FM and Z-AFM phases result from the competition between ferromagnetic or antiferromagnetic exchange interaction and easy-axis anisotropy. Z-FM phase is the ferromagnetic phase where spins orient along the $ \pm z $ axis. The spin structure factor $\bm{F_k}$ exhibits a peak at $\bm{k}=(0,0)$, which is the $\bm{\Gamma}$ point in the first Brillouin zone. Analogously, the Z-AFM phase is the antiferromagnetic phase where spins point along with the $ \pm z$ axis. The $\bm{F_k}$ exhibits peaks at $(\pm 2\pi/3, \pm 2\sqrt{3}\pi/3)$ and $ (\pm 4\pi/3,0)$.

Spiral phases occupy a large region in phase diagram without fields, which mostly result from the DM interaction. Spiral phases are coplanar states, where spins are all parallel at a plane and their direction rotates by a constant angle from one spin to a neighboring spin along the helical axis. 
When $\theta $ and $\Delta $ are small, the ferromagnetic exchange interaction is much more than the DM interaction, which leads to IC-Spiral phases in zero fields.  IC-Spiral phases are incommensurate spiral states. As $\theta $ increases, the DM interaction increasing, the emergence of $N^{*} \times 1$ spiral phases in phase diagram. $N^{*} \times 1$ spiral phases are different sizes of spiral states with $2N^{*}$ sites unit cell, where $N^{*}=2,3,4,5,6$ and $N^{*}$ decreased with $\theta$ increasing. 
Spins spiral in the $z-\bm{k}$ plane and the $N^{*} \times 1$ spiral phases are triple degeneracy because of the rotation symmetry of lattice. 
However, with the magnetic fields applied, the spiral phases could turn into IC-SkL phases rapidly, only the $3\times1$ and $4\times1$ spiral phases still stable with magnetic fields. In particular, the $2\times1$ spiral phase turn into the $2\times2$ FM star phase and the $\sqrt{3} \times \sqrt{3}$ vortex superlattice phase turn into the $\sqrt{3} \times \sqrt{3}$ z-vortex superlattice phase under magnetic fields, which preserve the rotation symmetry. 

Besides FM, AFM and spiral phases, we find rich varieties of topological spin textures in phase diagrams, such as SkL, VL, IC-SkL, FM star and Z-VL phases, which have significant potential for spintronics applications. Remarkably, these topological spin textures remain a large region in phase diagrams even with magnetic fields, which is important for spintronics applications. We describe the structures of these topological spin textures in details in the following and especially focus on these non-trivial phases only existing with magnetic fields. 

\subsection{Structures of topological spin textures}\label{sec-magnetic}
We describe both the real-space spin configurations and structure factors of all topological spin textures, where the structure factor is an important observation that can be used for the identification of magnetic phases experimentally.
Most of the phases in the phase diagram without fields are still stable when applying external magnetic fields. Besides SkL and VL topological phases, three field-induced topological phases have emerged, viz IC-SkL phase, FM star phase and Z-VL phase. We investigate all these topological spin textures in details as follows.

\subsubsection{$3 \times 3$ skyrmion superlattice and Incommensurate skyrmion superlattice}
\begin{figure} [ht!]  
	\begin{center}
		\includegraphics[width=0.98\linewidth]{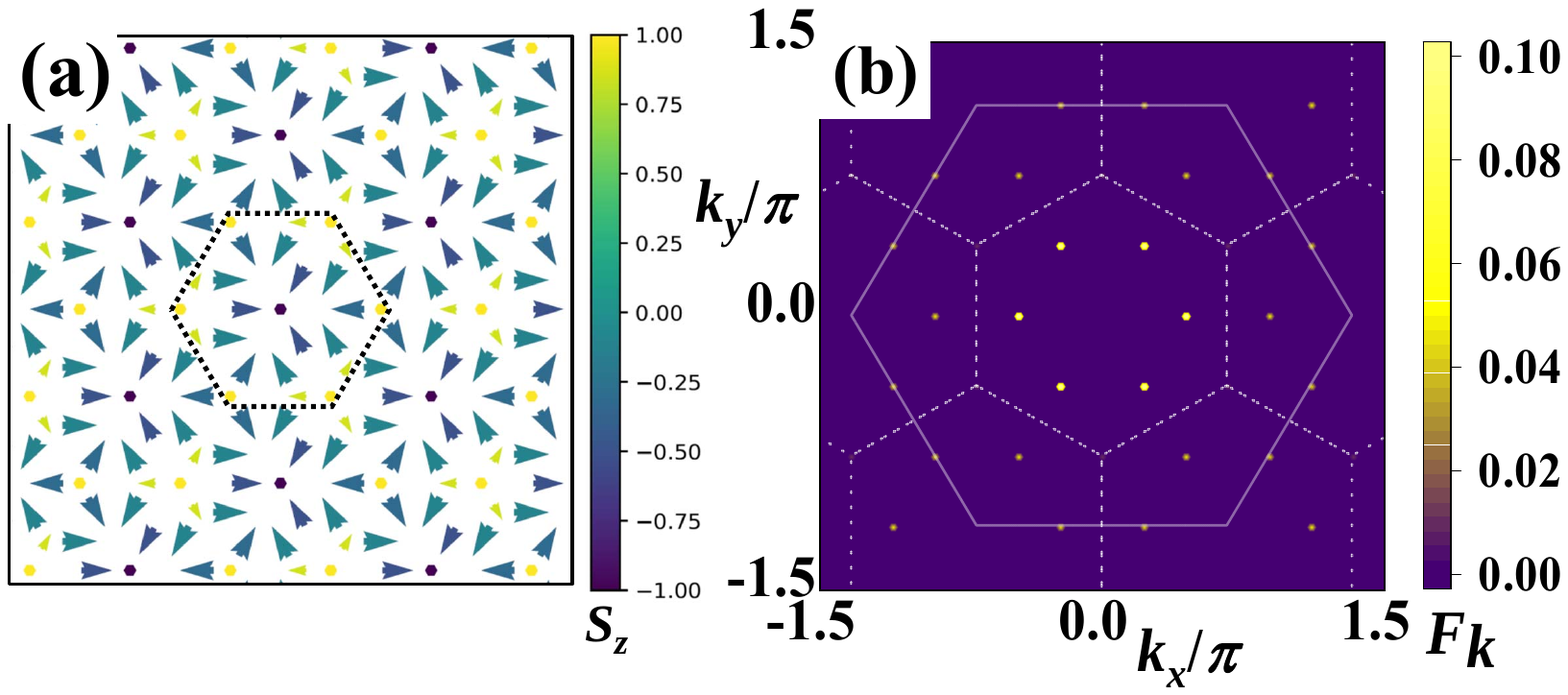}
	\end{center}
	\caption{  Structures of $3 \times 3$ SkL phase. (a) shows the fragment of real-space spin configurations in the $3 \times 3$ SkL phase, where the dashed line denote the magnetic unit cell. (b) shows the spin structure factor of $3 \times 3$ SkL phase, which has peaks at $(\pm 2\pi/9,\pm 2\sqrt{3}\pi/9)$ and $(\pm 4\pi/9,0)$, corresponding to $\bm{k}=\frac{2}{3}\mathbf{M}$ points.}
	\label{3-3-skl}
\end{figure}

The extremely strong competition between DM interaction and $A_c$ term leads to the $3 \times 3$ SkL phase. The SkL phase is a non-coplanar state, where the spins form an 18-site skyrmion structure. In Fig.~\ref{3-3-skl}(a), we give the real-space spin configuration of the SkL phase and the dashed line denotes the 18-site skyrmion magnetic unit cell. The elementary skyrmion contains a large hexagon with $3 \times 3$ unit cell of the honeycomb lattice. In the hexagon of $3 \times 3$ unit cell, a spin with negative $S^z$ magnetization at the core, spins from the core lie in the $xy$-plane rotating $2\pi$ and six outermost spins with positive $S^z$ magnetization. The structure factor of SkL phase has peaks at $\bm{k}=\frac{2}{3}\mathbf{M}$ points shown in Fig.~\ref{3-3-skl}(b).
We calculated the topological charge for the $3 \times 3$ SkL phase by the definition of topological charge shown in Fig.~\ref{zero}(c). The elementary skyrmion was marked with the dashed line in Fig.~\ref{3-3-skl}(a), where the topological charge $|Q|=1$ for each 18-site skyrmion. 
\begin{figure*} [ht!]  
	\begin{center}
		\includegraphics[width=0.83\linewidth]{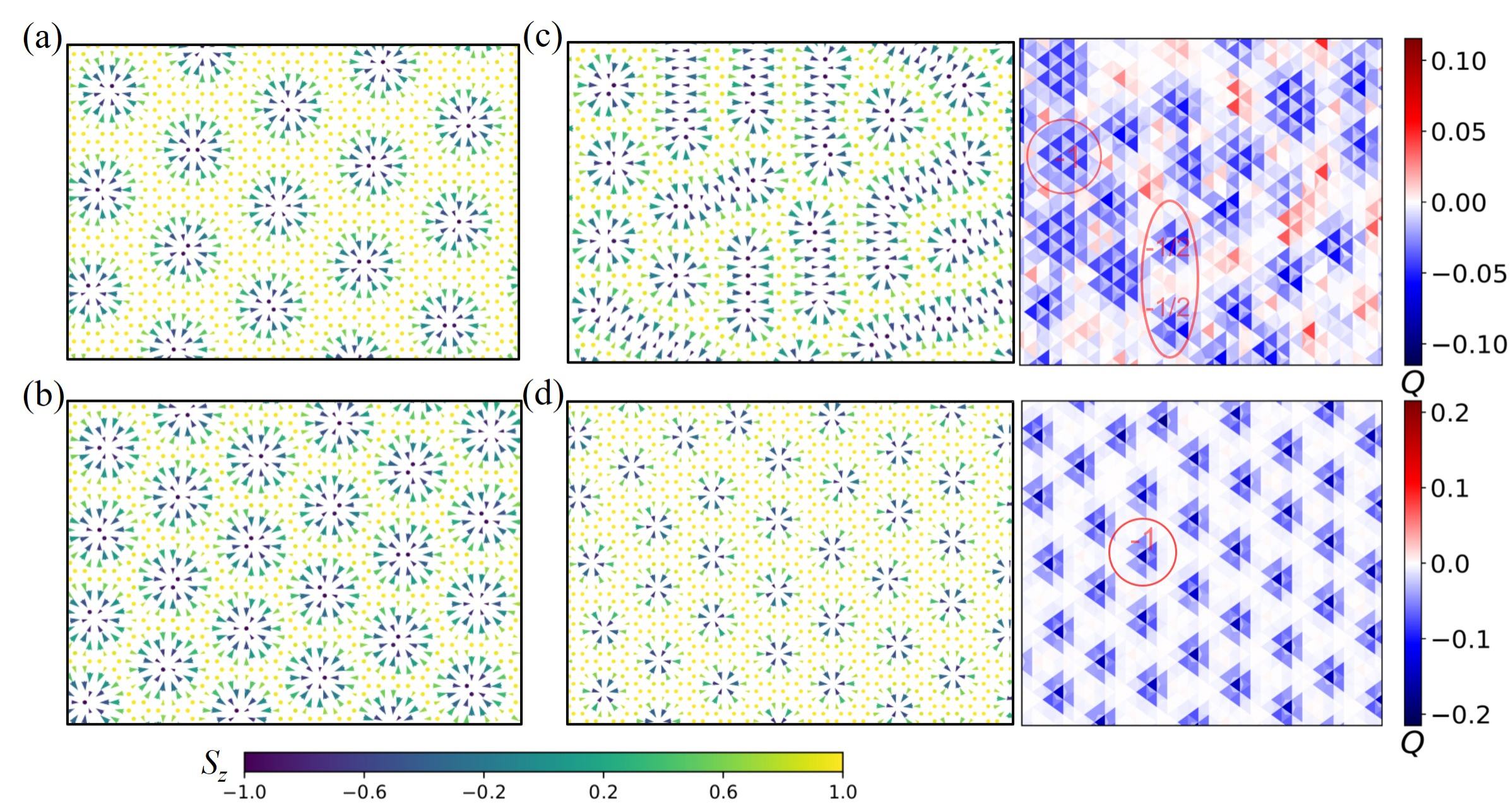}
	\end{center}
	\caption{  Structures of IC-SkL phase. (a) Fragment of IC-SkL state at $\Delta=0.2$, $H/|J|=0.2$, $\theta=0.10 \pi$. (b) Fragment of IC-SkL state at $\Delta=0.2$, $H/|J|=0.2$, $\theta=0.13 \pi$. (c) Fragment of IC-SkL state at $\Delta=0.5$, $\theta=0.15 \pi$, $H/|J|=0.2$ including mixed skyrmion-bimeron states and topological charge density shown in the right. Bimerons are marked with red ovals and skyrmions are mark with red cycles. (d) Fragment of IC-SkL state at $\Delta=0.5$, $\theta=0.15 \pi$, $H/|J|=0.5$ including skyrmions and topological charge density shown in the right. Skyrmions are mark with red cycles. }
	\label{phases}
\end{figure*}
\begin{figure*} [ht!]  
	\begin{center}
		\includegraphics[width=0.24\linewidth]{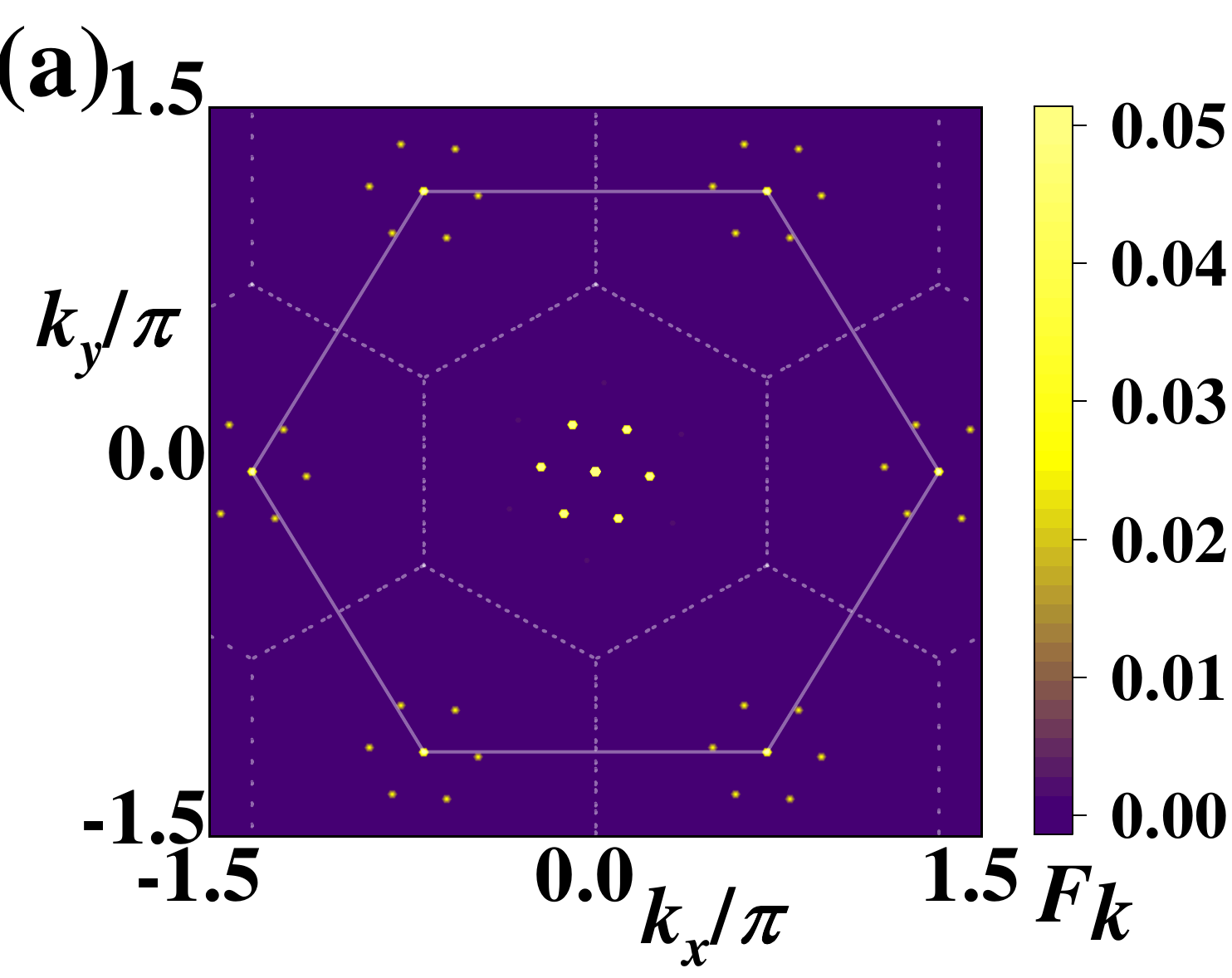}
		\includegraphics[width=0.24\linewidth]{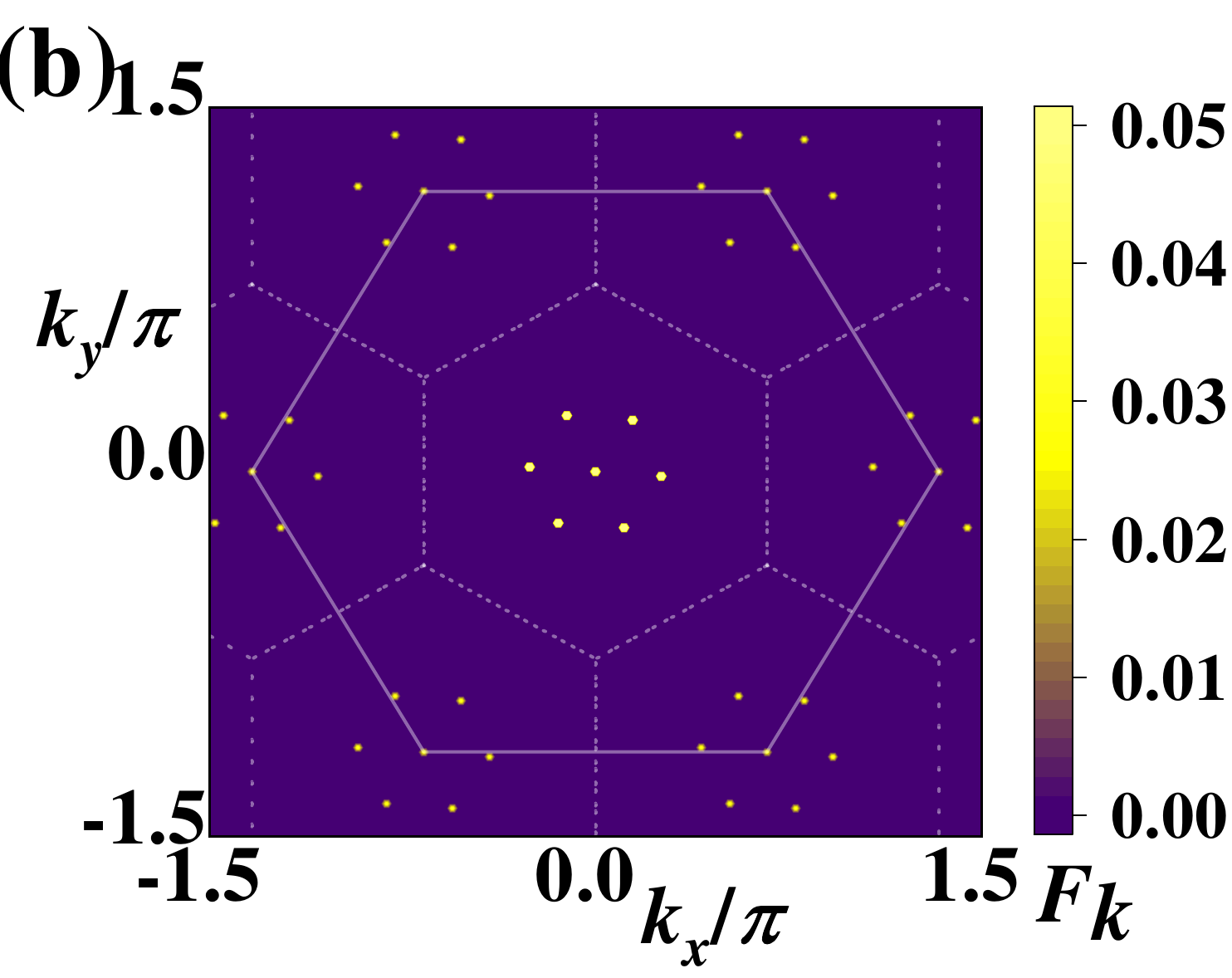}
		\includegraphics[width=0.24\linewidth]{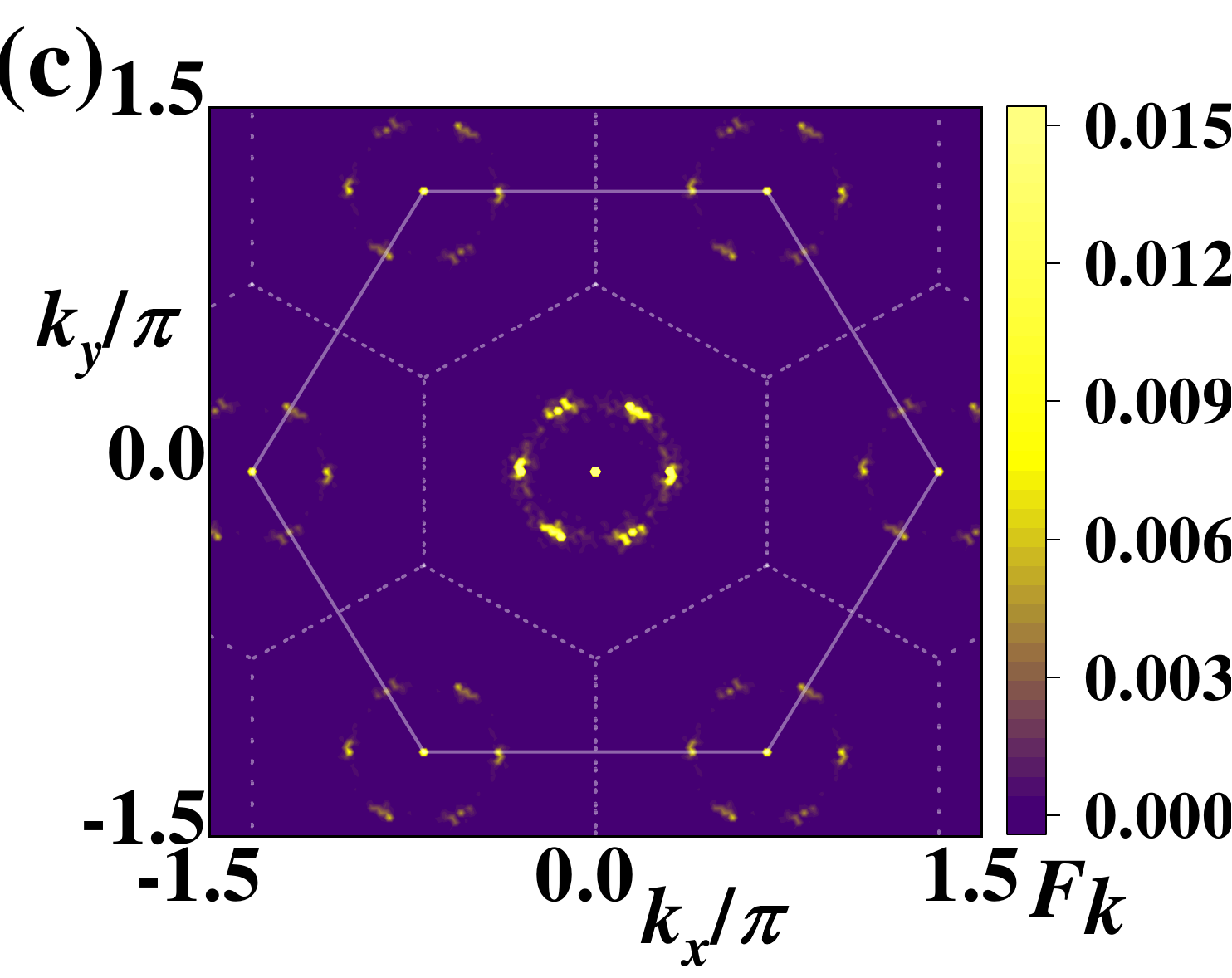}
		\includegraphics[width=0.24\linewidth]{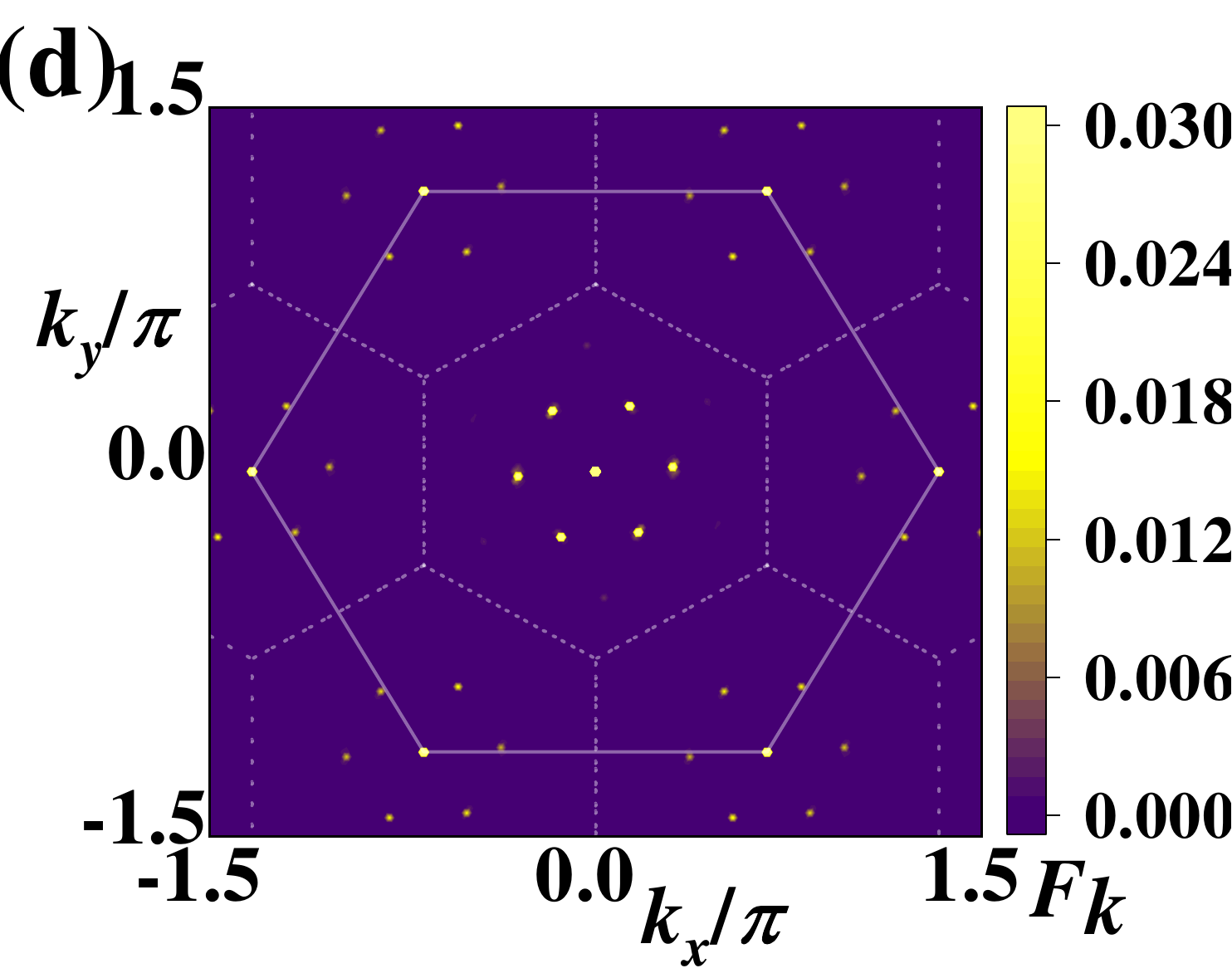}
	\end{center}
	\caption{  The spin structure factors in IC-SkL phase. These results correspond to Fig.~\ref{phases}(a)-(d), respectively. (a), (b) and (d) show the structure factors have different peaks in different IC-SkL states because of the change in skyrmion number. (c) shows the structure factor of mixed skyrmion-bimeron states has spread peaks because of the existence of bimerons.}
	\label{phases-sk}
\end{figure*}
\begin{figure} [ht!]  
	\begin{center}
		\includegraphics[width=0.88\linewidth]{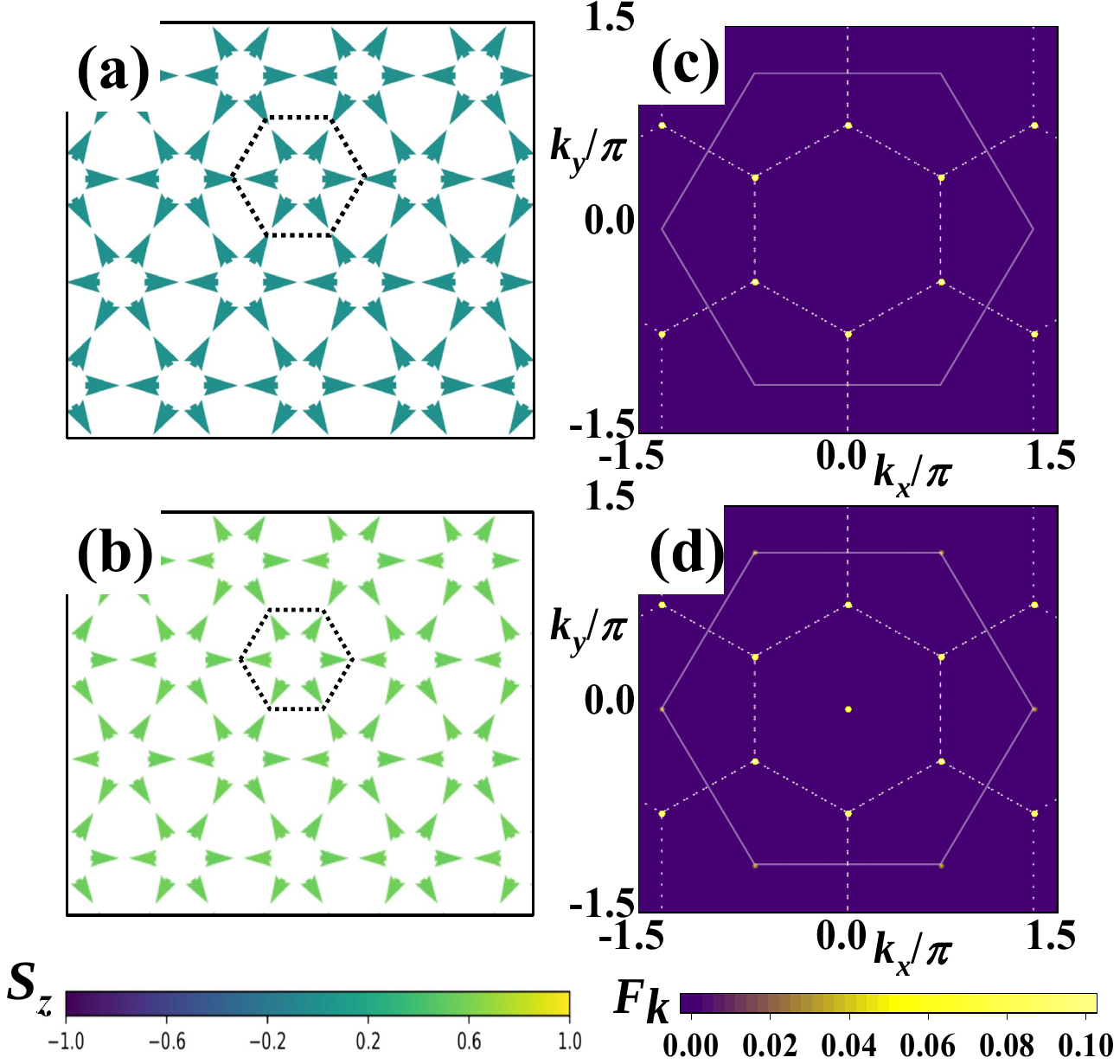}
	\end{center}
	\caption{  Structures of $\sqrt{3} \times \sqrt{3}$ VL and $\sqrt{3} \times \sqrt{3}$ Z-VL phases. (a)-(b) show the fragment of real-space spin configurations in VL and  Z-VL phases, where the dashed line denote the magnetic unit cell. (c)shows the spin structure factor of VL phase, which has peaks at $\mathbf{K}$ points. (d) shows the spin structure factor of Z-VL phase, which has peaks at $\mathbf{K}$ points and $\mathbf{\Gamma}$ points.}
	\label{vl-z-vl}
\end{figure}
\begin{figure} [ht!]  
	\begin{center}
		\includegraphics[width=0.98\linewidth]{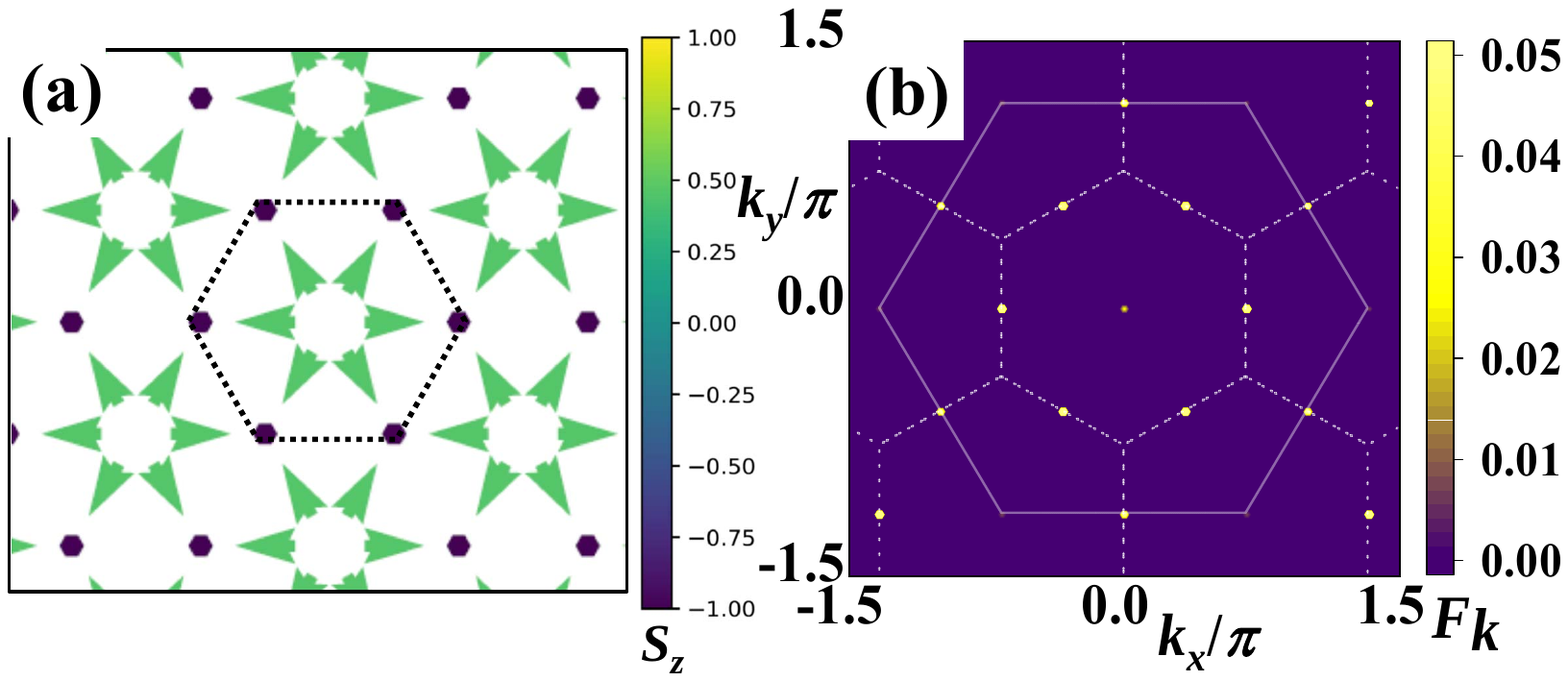}
	\end{center}
	\caption{  Structures of $2 \times 2$ FM star phase. (a) shows the fragment of real-space spin configurations in $2 \times 2$ FM star phase, where the dashed line denote the magnetic unit cell. (b) shows the spin structure factor of $2 \times 2$ FM star phase, which has peaks at $\mathbf{M}$ points and $\mathbf{\Gamma}$ points.}
	\label{fm-star}
\end{figure}

There are similar IC-SKL phases exist in the phase diagrams with magnetic fields. The IC-Spiral states and larger size spiral states are not stable when applying the magnetic fields. With the magnetic fields increasing, the IC-Spiral states and large-size spiral states are superseded by IC-SkL phases rapidly. The IC-SkL phases are complicated states including the different sizes of skyrmion superlattice and mixed skyrmion-bimeron states, which result from the extremely strong competition among the ferromagnetic exchange interaction, DM interaction and the external magnetic fields. 

IC-SkL phases are non-coplanar topological states, including non-coplanar skyrmion states and mixed skyrmion-bimeron states. In Fig.~\ref{phases}(a)-(d), we give four spin configurations of IC-SkL states corresponding to different parameters. In Fig.~\ref{phases}(a)-(b), we give the spin configurations of IC-SkL states by tuning the parameter $\theta$.  From Fig.~\ref{phases}(a) to Fig.~\ref{phases}(b), we fixed the value of $\Delta=0.2$ and $H/|J|=0.2$, then set up the $\theta$ equal to $0.10\pi$ and $0.13\pi$. The increasing of the DM interaction with tuning $\theta$ leads to the size of skyrmions decreasing and the number of skyrmions increasing, corresponding to the order parameter topological charge in Fig.~\ref{field}(c). As shown in Fig.~\ref{phases-sk}(a)-(b), the region occupied by the peaks in the structure factors also indicate the size of skyrmions decreasing.   

In Fig.~\ref{phases}(c)-(d), we give the spin configurations and corresponding topological charge density of IC-SkL states by tuning the external magnetic fields $H/|J|$. From Fig.~\ref{phases}(c) to Fig.~\ref{phases}(d), we fixed the value of $\Delta=0.5$ and $\theta=0.15\pi$, then set up the $H/|J|$ equal to $0.2$ and $0.5$. As the increasing of the magnetic fields with parameter $H/|J|$, bimerons states contracting and the mixed skyrmion-bimeron states are superseded by the skyrmion superlattice gradually. In the spin configuration of mixed states (Fig.~\ref{phases}(c)), we could find skyrmions and bimerons, which consist of two merons carrying the topological charge $Q=-1/2$ and connected by a spiral domain with zero topological charge and are relatively stable in low fields.  In Fig.~\ref{phases-sk}(c)-(d), we give the structure factors of mixed skyrmion-bimeron states and skyrmion lattice, where the structure factor of mixed states has spread peaks because of the existence of bimerons. 

\subsubsection{$\sqrt{3} \times \sqrt{3}$ vortex superlattice and $\sqrt{3} \times \sqrt{3}$ z-vortex superlattice}

At $\theta$ is close to $0.5\pi$, the DM interaction is almost to zero and the $A_c$ term reaches its maximum. When $\Delta $ is close to $0$, the spins coplanar in $x y$ plane because of the $A_c$ term, which lead to another topological spin texture, $\sqrt{3} \times \sqrt{3}$ VL phase.
The VL phase is a coplanar state and spins in $xy$-plane form a $\sqrt{3} \times \sqrt{3}$ unit cell vortex, where each vortex contains 6 sites. In Fig.~\ref{vl-z-vl}(a), We give the real-space spin configuration of the VL phase and the dashed line denotes the 6-site vortex magnetic unit cell. The elementary vortex contains a hexagon with a 6-site vortex and in each vortex, the spins in $x y$-plane wind anti-clockwise $2\pi$ around each hexagon. 
The structure factor of VL phase has peaks at $(\pm 2\pi/3,\pm 2\sqrt{3}\pi/9)$ and $(0,\pm 4\sqrt{3}\pi/9)$ shown in Fig.~\ref{vl-z-vl}(c).

We get the $\sqrt{3} \times \sqrt{3}$ Z-VL phase in the region between VL phase and Z-AFM phase with magnetic fields, which mostly result from the competition between the $A_c$ term and the external out-of-plane magnetic fields. The $A_c$ term is favorable to form the in-plane vortex superlattice state. Whereas, when applying the magnetic fields, the out-of-plane components of spins show ferromagnetic order. The Z-VL phase is also a non-coplanar state, where the in-plane components of spins form a $\sqrt{3} \times \sqrt{3}$ unit cell vortex. In Fig.~\ref{vl-z-vl}(b), We give the real-space spin configuration of the Z-VL phase and the dashed line denotes a 6-site magnetic unit cell. 
However, the out-of-plane components of spins show ferromagnetic order. The structure factor of  Z-VL phase is similar to VL phase. There has an extra peak at $(0,0)$, which is $\bm{\Gamma}$ point, corresponding to the ferromagnetic order of out-of-plane components shown in Fig.~\ref{vl-z-vl}(d).

\subsubsection{$2 \times 2$ FM star}
We get the $2 \times 2$ FM star state in the region of $2 \times 1$ spiral phase with applied external magnetic fields. The FM star state is another non-coplanar state, where the spins form an 8-site magnetic unit cell with vortex-like structure. The magnetic fields lead to the transformation from the $2 \times 1$ spiral phase to the $2 \times 2$ FM star. As in zero fields, the $2 \times 1$ spiral phase breaks the rotation symmetry, whereas the FM star phase preserves the rotation symmetry. The FM star phase is favored by the competition between $A_c$ term, DM interaction, and magnetic fields.  In Fig.~\ref{fm-star}(a), we give the real-space spin configuration of the FM star phase and the dashed line denotes an 8-site magnetic unit cell. The whole FM star contains a large hexagon with $2 \times 2$ unit cell of the honeycomb lattice. In the hexagon of an 8-site magnetic unit cell, the in-plane spin component rotating $2\pi$ in the core hexagon and six spins in the vertices of the outer hexagon are aligned in a parallel fashion. 

To characterize and identify the $2 \times 2$ FM star phase, we also calculate the structure factor. In Fig.~\ref{fm-star}(b), we give the structure factor of FM star phase, which has peaks at $(\pm \pi/3,\pm \sqrt{3}\pi/3)$, $(\pm 2\pi/3,0)$ and an extra peak at $(0,0)$. These peaks are $\mathbf{M}$ points of the first Brillouin zone and the peak at $\bm{\Gamma}$ point corresponding to the ferromagnetic order of six outermost spins. The $2 \times 2$ FM star phase remains stable in a large parameter region with magnetic fields, which could be expected to be found in future experimental work.

\section{Discussion and Conclusion}\label{sec-discuss}
In this paper, we find various topological spin textures. Remarkably, these topological spin textures remain a large region in phase diagrams even with magnetic fields, which have significant potential for spintronics applications and could further understand the effect of magnetic fields on topological spin textures. These novel topological spin textures could be reproduced in both real chiral magnets and ultra-cold atoms systems. We hope that our results could be achieved in future experimental work.  

In conclusion, we study the rotated Heisenberg model on the honeycomb lattice via Monte Carlo simulations, which could describe the low-energy behavior of a bosonic model with SOC in the deep Mott insulating region or two-dimensional chiral magnets with strong SOC. We obtain the classical phase diagrams with/without fields and mainly focus on phase diagrams with fields, especially on the non-trivial phases only existing with fields. We found $N^*\times1$ spiral phases, IC-Spiral phase, SkL phase and VL phase in zero fields. As the magnetic fields were applied, we found IC-SkL phases, where the number of skyrmions changes with $\theta$ and mixed skyrmion-bimeron states appear at $\Delta=0.5$, $\theta=0.15 \pi$, $H/|J|=0.2$, FM star phase and Z-VL phase.
We describe both spin configurations and spin structure factors of topological spin textures in details, where the structure factor of mixed skyrmion-bimeron states has spread peaks compared with SkL states. We also found the structure factor of FM star phase has peaks at $\mathbf{M}$ points and the structure factor of VL phase has peaks at $\mathbf{K}$ points. Overall, our results of these topological spin textures could have significant potential for the identification of topological phases experimentally.

\section{Acknowledgments}
We thank Ji-Ze Zhao for the useful discussion. This work was supported by the National Natural Science
Foundation of China (Grants No. 11774300).

\end{document}